\documentclass{PoS}

\def\simle{\mathrel{\rlap{\raise 0.511ex \hbox{$<$}}{\lower 0.511ex
 \hbox{$\sim$}}}}
\newcommand{\bea}{\begin{eqnarray}}
\newcommand{\eea}{\end{eqnarray}}
\newcommand{\be}{\begin{equation}}
\newcommand{\ee}{\end{equation}}
\newcommand{\nn}{\nonumber}
\newcommand{\gev}{{\rm~GeV}}
\newcommand{\mev}{{\rm~MeV}}

\newcommand{\msb}{\overline{\rm{MS}}}
\newcommand{\mbar}{\overline{m}}

\title{Quark masses with $N_f=2$ twisted mass lattice QCD}

\ShortTitle{Quark masses with $N_f=2$ tmQCD}

\author{B.~Blossier\\
Lab. de Phys. Th\'eorique, Univ. de Paris XI, Centre d'Orsay, 91405 Orsay-Cedex, France}

\author{P.~Dimopoulos\\
Dipartimento di Fisica, Universit\`a di Roma ``Sapienza'', I-00185 Roma, Italy}

\author{R.~Frezzotti, G.~C.~Rossi\\
Dip. di Fisica, Univ. di Roma Tor Vergata, and INFN, Via della Ricerca Scientifica, I-00133 Roma, Italy}

\author{V.~Lubicz, C.~Tarantino\\
Dip. di Fisica, Universit{\`a} Roma Tre, and INFN, Via della Vasca Navale
84, I-00146 Roma, Italy}

\author{M.~Petschlies\\
Institut f\"ur Elementarteilchenphysik, Fachbereich Physik,
\\ Humboldt Universit\"at zu Berlin, D-12489, Berlin, Germany}

\author{\speaker{F.~Sanfilippo}\\
Dipartimento di Fisica, Universit\`a di Roma ``Sapienza'', and INFN, I-00185 Roma, Italy}

\author{S.~Simula\\
INFN, Sez. di Roma
Tre, Via della Vasca Navale 84, I-00146 Roma, Italy}

\author{for the European Twisted Mass Collaboration (ETMC)}

\abstract{We present the results of the recent high precision lattice calculation of the average up/down, strange
and charm quark masses performed by ETMC with $N_f=2$ twisted mass Wilson fermions. The
analysis includes data at four values of the lattice spacing and pion masses as
low as $\simeq 270$ MeV, allowing for accurate continuum limit and chiral extrapolation.
The strange and charm masses are extracted by using several methods, based on
different observables: the kaon and the $\eta_s$ meson for the strange
quark and the $D$, $D_s$ and $\eta_c$ mesons for the charm. The quark mass
renormalization is carried out non-perturbatively using the RI-MOM method. The
results for the quark masses in the $\msb$ scheme read: $\mbar_{ud}(2\,\gev)=
3.6(2)\mev$, $\mbar_s(2\gev)=95(6)\mev$ and $\mbar_c(\mbar_c)=1.28(4)\gev$.
We have also obtained the ratios $m_s/m_{ud}=27.3(9)$ and $m_c/m_s=12.0(3)$.
Moreover, we provide the updated result for the bottom quark mass, $\mbar_b(\mbar_b)=4.3(2)\,\gev$, obtained using the method presented in~\cite{Blossier:2009hg}.}

\FullConference{The XXVIII International Symposium on Lattice Field Theory\\
		June 14-19, 2010\\
		Villasimius, Sardinia Italy}

\begin{document}

\section{Introduction}
We present the recent accurate determination~\cite{lavoro} of the average up/down,
strange and charm quark masses performed by ETMC with $N_f=2$ maximally twisted mass Wilson fermions. The high precision of this analysis is mainly due to the extrapolation of the
lattice results to the continuum limit, based on data at four values of the
lattice spacing ($a \simeq$ 0.098, 0.085, 0.067, 0.054 fm), to the well controlled chiral extrapolation, which uses
simulated pion masses down to $M_\pi \simeq 270 \mev$, and to the use of the
non-perturbative renormalization constants calculated
in~\cite{Constantinou:2010gr}. The only systematic uncertainty which is not
accounted for is the one due to the missing strange and
charm quark vacuum polarization effects.
However, a comparison of $N_f=2$ results for the up/down and
strange quark masses to already existing results from $N_f=2+1$ quark flavor
simulations~\cite{Scholz:2009yz} indicates that, for these observables, the error
due to the partial quenching of the strange quark is smaller at present than
other systematic uncertainties. 
The same conclusion is expected to be valid for the effects of the strange and charm partial quenching in the determination of the charm quark mass.
In this respect we mention that simulations with $N_f=2+1+1$ dynamical flavors are already being performed by ETMC and preliminary results for several flavor physics observables have been recently presented~\cite{Baron:2010bv,Baron:2010th}. For more details on the ensembles of $N_f=2$ gauge configurations used in the analysis and the values of the simulated light, strange and charm quark masses we refer to~\cite{lavoro}.

The calculation of the averaged up/down quark mass, based on the study of
the pion mass and decay constant, has closely followed the strategy
of~\cite{Baron:2009wt}. At variance with the latter, however, here and in~\cite{lavoro} data at four values of the lattice spacing have been used.
For the strange quark mass, the main improvement
with respect to our previous work~\cite{Blossier:2007vv}, which used data at
a single lattice spacing only, is the continuum limit. Moreover, the
chiral extrapolation has been performed by using either SU(2)- or SU(3)-Chiral
Perturbation Theory (ChPT). In order to extract the
strange quark mass we have used both the kaon mass and the mass of the (unphysical) $\eta_s$
meson composed of two degenerate valence strange quarks.
For the charm quark mass, similarly to the strange quark, we have used
several experimental inputs to extract its value: the mass of the $D$, $D_s$ and
$\eta_c$ mesons.
 
The results that we have obtained for the quark masses are, in the $\msb$ scheme, 
\bea
&& \mbar_{ud}(2\gev) = 3.6(2)\mev\,,\nn\\
&&\mbar_s(2\gev) = 95(6)\mev\,,\nn\\
&&\mbar_c(\mbar_c) = 1.28(4)\gev\, .
\label{eq:res1}
\eea
We have also obtained for the ratios of quark masses the values 
\be
m_s/m_{ud}=27.3(9) \qquad {\rm and } \qquad m_c/m_s=12.0(3) \ , 
\ee
which are independent of both the renormalization scheme and scale.

Finally, we take the opportunity of these proceedings to present the updated result for the bottom quark mass which has been  obtained employing the method discussed in~\cite{Blossier:2009hg}.
The updated value is
\be
\mbar_b(\mbar_b)=4.3(2)\,\gev \ .
\ee

\section{Up/down quark mass}
We have studied the dependence of the pion mass and decay constant on the
renormalized quark mass, by using the predictions based on NLO ChPT and the Symanzik expansion up to $\mathcal{O}(a^2)$.\footnote{Here and in the following sections we refer to~\cite{lavoro} for the expressions of the fitting functions.}
They include discretization terms of $\mathcal{O}(a^2 m_l \log(m_l))$, which receive a contribution from the $\mathcal{O}(a^2)$ splitting between the neutral and charged pion mass~\cite{Bar:2010jk} occurring with twisted mass fermions. The impact of this correction on the final result for the light quark mass is at the level of the fitting error.

Lattice results for pion masses and decay constants have been corrected for finite size effects (FSE) evaluated using the resummed L\"uscher formulae.
The effect of the $\mathcal{O}(a^2)$ isospin breaking has been taken into account also in these corrections~\cite{Colangelo:2010cu}.
On our pion data, FSE vary between $0.2$\% and 2\%, depending on the simulated mass and volume.
The inclusion of the pion mass splitting in the FSE induces an effect at the level of one third of the statistical error for our lightest pion mass at $\beta=3.9$ on the smaller volume, and even smaller in the other cases.

The value of the physical up/down quark mass is extracted from the ratio
$m_{\pi}^2/f_\pi^2$ using as an input the experimental value of the latter
ratio.
In order to estimate the systematic uncertainty due to discretization effects we have performed both a fit without the logarithmic discretization terms, and a fit without all $\mathcal{O}(a^2)$ corrections. Both these ans\"atze turn out be compatible with the lattice data.
We find that the result for the up/down quark mass decreases by approximately 2\% and increases of about 6\% in the two cases respectively, so that we estimate an overall uncertainty due to residual discretization effects of $\pm 4$\%.
For estimating the systematic uncertainty due to the chiral extrapolation we have also considered a fit including a NNLO local contribution proportional to the light quark mass square. In this case we are not able to determine all the fitting parameters and we are thus forced to introduce, on the additional LECs, priors as in~\cite{Baron:2009wt}. In this way we find that the result for $m_{ud}$ increases by $6$\%.
We have also included in the final result a 2\% systematic uncertainty coming from the perturbative conversion of the quark mass renormalization constant from the RI-MOM to the $\msb$ scheme.
This uncertainty has been conservatively estimated by assuming the unknown $\mathcal{O}(\alpha_s^4)$ term to be as large as the $\mathcal{O}(\alpha_s^3)$ one, evaluated at the renormalization scale $\mu\simeq 3\,\gev$, which is the typical scale of the non-perturbative RI-MOM calculation in our simulation~\cite{Constantinou:2010gr}. 
Adding in quadrature the three systematic errors discussed above we have obtained
\be
\mbar_{ud}(2\,\gev) = 3.6(1)(2) \mev\,= 3.6(2) \mev\,.
\label{eq:rismud}
\ee

\section{Strange quark mass}
In this section, we first present the determination of the strange quark mass
based on the study of the kaon meson mass and then the alternative method based
on the study of the $\eta_s$ meson.

In order to better discriminate the strange quark mass dependence of the kaon
masses on other dependencies, we have firstly interpolated, by using quadratic splines, the lattice data to three reference values of the strange quark
mass, chosen to be equal at the four lattice spacings: $\mbar_s^{ref} = \{ 80\,,\ 95\,,\ 110\} \mev$.
Then, at fixed reference
strange mass, we have simultaneously studied the kaon mass dependence on the up/down
quark mass and on discretization effects, thus performing a combined chiral
and continuum extrapolation. In this step, we have considered
chiral fits based either on SU(2)-ChPT~\cite{Allton:2008pn,Roessl:1999iu} or partially
quenched SU(3)-ChPT~\cite{Sharpe:1997by}. Finally, we have studied the kaon mass dependence on the strange quark mass, and determined the value of the physical strange quark mass using the experimental value of $m_K$.
\begin{figure}[tb]
\includegraphics[width=0.38\textwidth,angle=270]{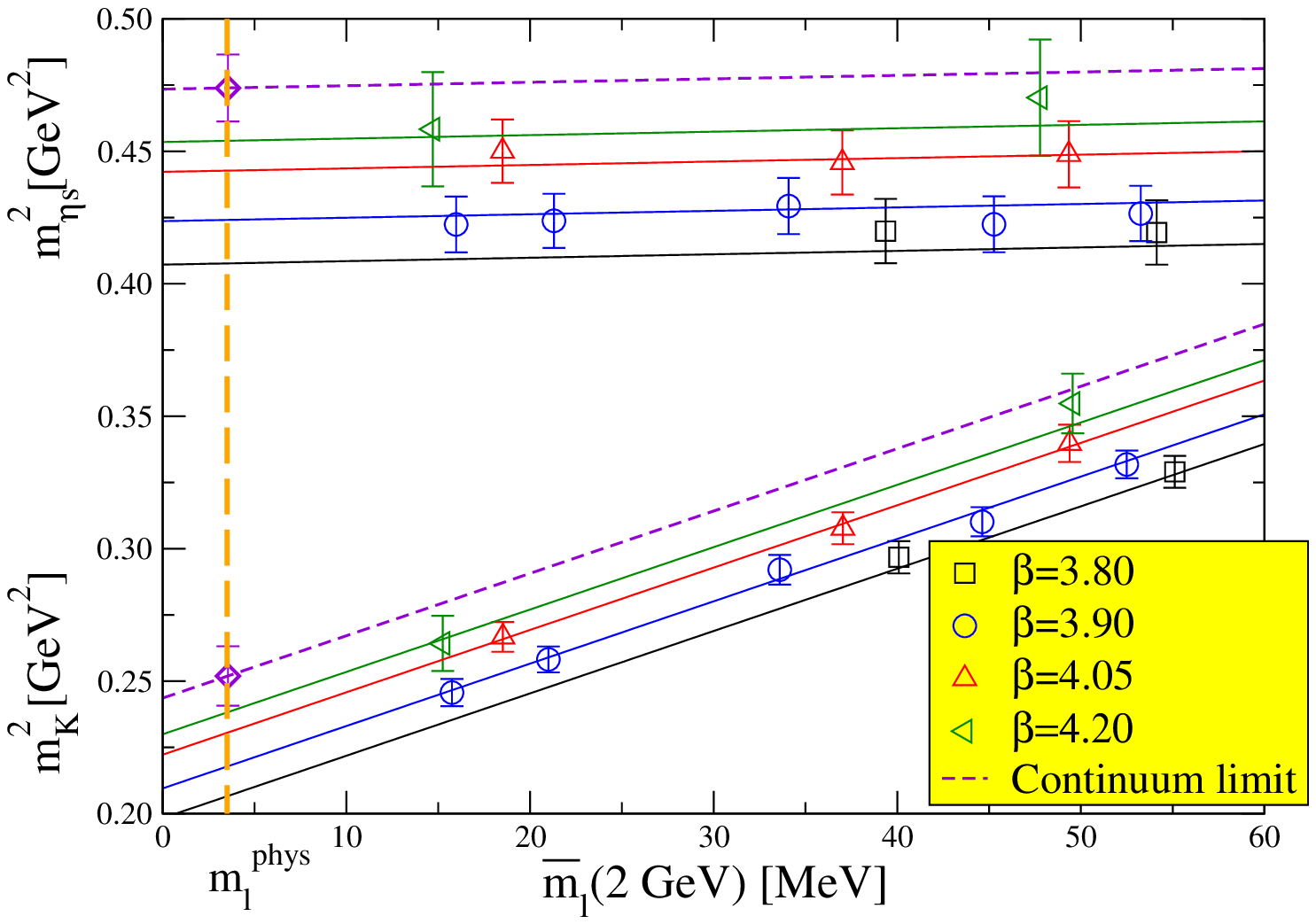}
\hspace{-0.8cm}
\includegraphics[width=0.38\textwidth,angle=270]{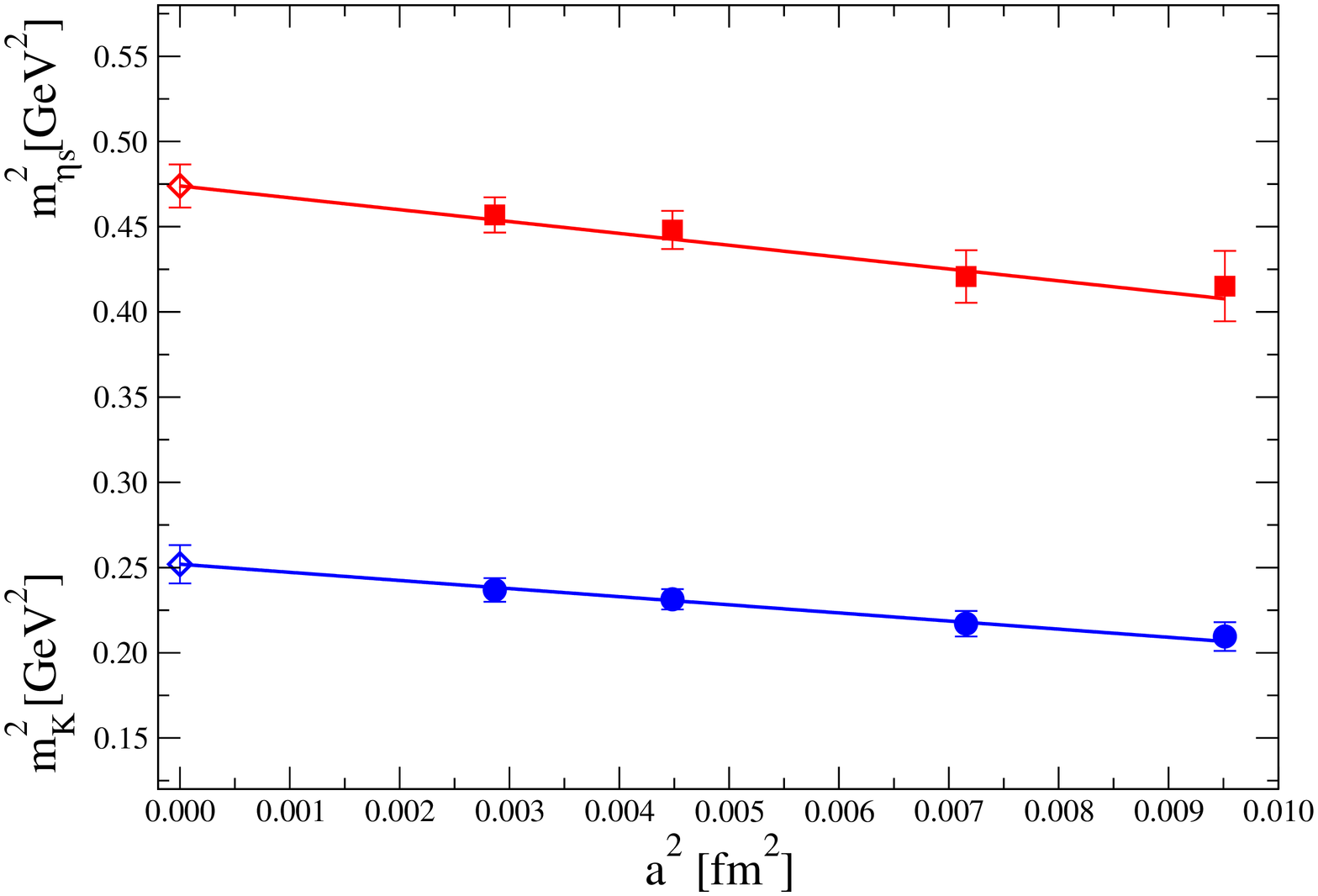}
\caption{\sl Left: Dependence of $m_K^2$ and $m_{\eta_s}^2$ on the renormalized light quark mass, for a
fixed reference strange quark mass ($\mbar_s^{ref}=95\,\mev$) and at the four lattice spacings.
Right: Dependence of $m_K^2$ and $m_{\eta_s}^2$ on the squared lattice spacing, for $\mbar_s^{ref}=95\,\mev$ and at the physical up/down mass.
\label{fig:mKmss1}}
\end{figure}
\begin{figure}[tb]
\begin{center}
\includegraphics[width=0.4\textwidth,angle=270]{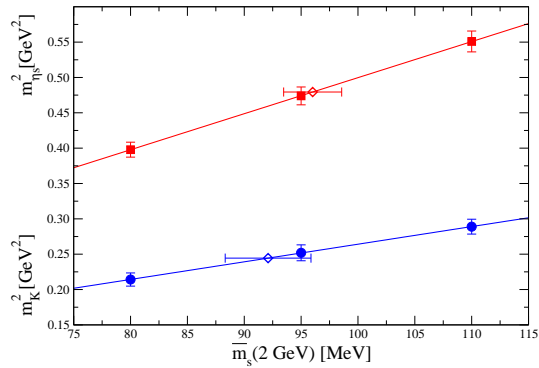} 
\caption{\sl Dependence of $m_K^2$ and $m_{\eta_s}^2$, in the continuum limit and at the physical up/down mass, on the strange quark mass. The strange mass results are also shown (empty diamonds).
\label{fig:mKmss2}}
\end{center}
\end{figure}
In fig.~\ref{fig:mKmss1} we show the combined chiral/continuum fit based on SU(2)-ChPT, for a
fixed reference value of the strange quark mass, as a function of the light quark mass (left) and of the squared lattice spacing (right). In fig.~\ref{fig:mKmss2} the dependence on the strange quark mass is shown, for the SU(2) analysis. The dependencies are shown for the kaon squared mass as well as for the $\eta_s$ squared mass discussed hereafter.

As an alternative way to determine the strange quark mass we have studied the
dependence on $m_s$ of a meson made up of two strange valence
quarks~\cite{Davies:2009ih}. The advantage of this approach is that the mass of 
this unphysical meson, denoted as $\eta_s$, is only sensitive to the up/down quark mass through sea quark effects, and thus requires only a very smooth chiral extrapolation. 
The price to pay is the need for an additional chiral fit to determine the $\eta_s$ mass at the physical point. In order to relate the mass of the $\eta_s$ meson to the physically observable $m_\pi$ and $m_K$, we have studied its dependence on the kaon and
pion masses for different values of the simulated light and strange quark
masses.
We investigated functional forms based on either SU(2)- or SU(3)-ChPT. The two fits yield very close results for the $\eta_s$ meson mass and we quote as final estimate $m_{\eta_s}=690(3)$ MeV (to be compared to the LO SU(3) prediction $m_{\eta_s}=(2\,m_K^2-m_\pi^2)^{1/2}=686$ MeV). 

Once the mass of the $\eta_s$ meson has been determined, the strange
quark mass can be extracted by following the very same procedure described for
the case of the kaon mass.

The difference between the determinations based on the $K$ and $\eta_s$ mesons is about 3\%.
The results obtained from either the SU(2) or the SU(3) fits are practically the same in the analysis based on the $\eta_s$ and differ by approximately 3\% in the kaon case.
In order to evaluate the uncertainty of the continuum extrapolation we have excluded from these fits the data from the coarser lattice, finding a variation of the results of approximately 2\%, with the fitting error approximately unchanged.
The different fits considered for the determination of the up/down mass and of the lattice spacing affect the determination of the strange mass at the level of 3\%.
Finally, we have included an uncertainty of 2\% related to the truncation of the perturbative expansion in the conversion from the RI-MOM to the $\msb$ scheme.
Combining all these uncertainties in quadrature, we quote as our final
estimate of the strange quark mass in the $\msb$ scheme
\be
\mbar_s(2\,\gev)=95(2)(6)\,\mev=95(6)\,\mev\,.
\label{eq:risms}
\ee
Using the determinations of both the strange and light quark masses, we have also obtained a prediction for the ratio $m_s/m_{ud}$, which is both a scheme and scale independent quantity: 
\be
m_s/m_{ud}=27.3(5)(7)=27.3(9) \ .
\ee

\section{Charm quark mass}
The determination of the charm quark mass follows, quite closely, the strategy
adopted in the determination of the strange quark mass discussed in the
previous section. In this case, we have used as experimental input the masses of the 
$D$, $D_s$ and $\eta_c$ mesons. 

\begin{figure}[tb]
\includegraphics[width=0.38\textwidth,angle=270]{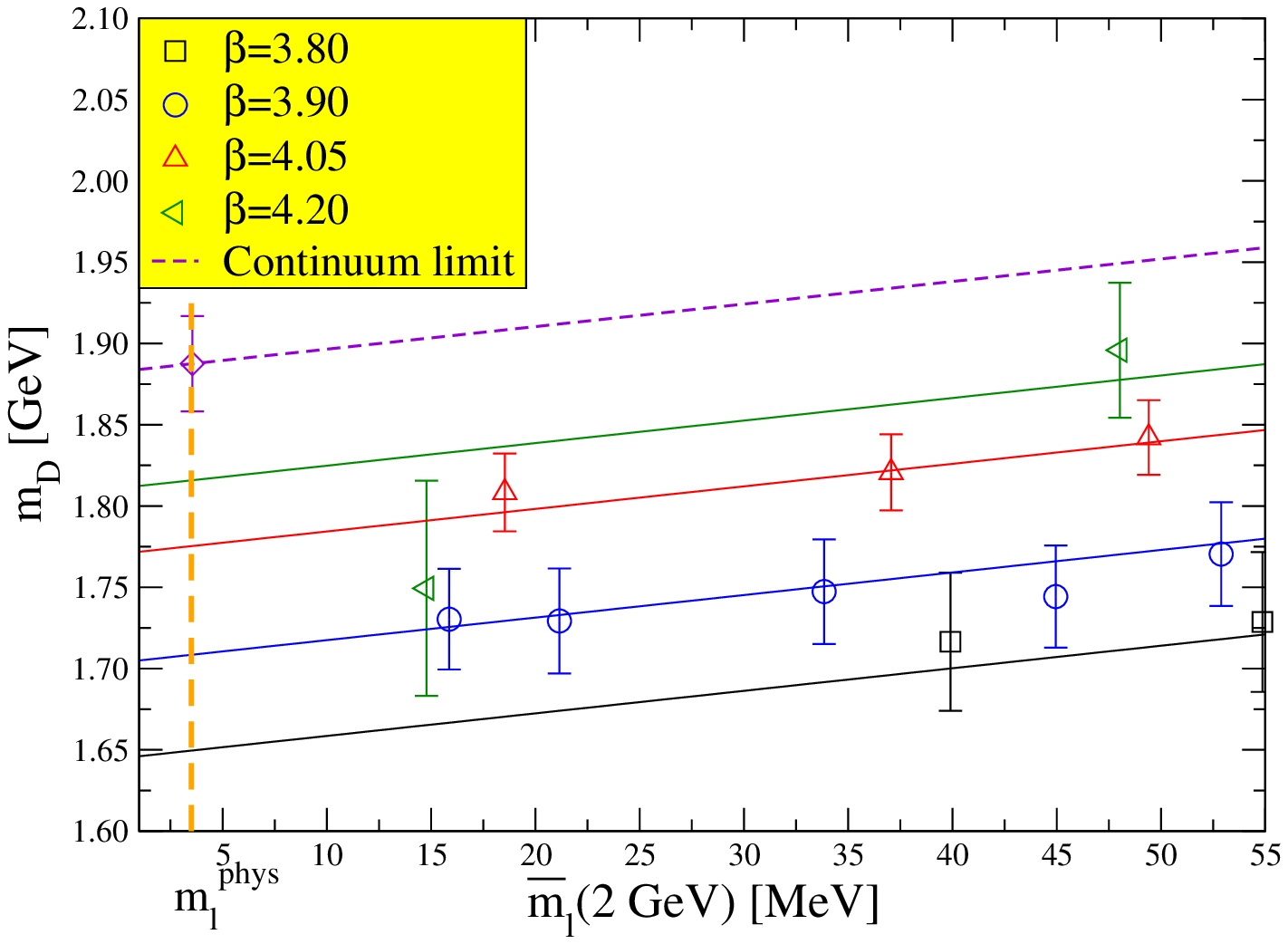}
\hspace{-0.8cm}
\includegraphics[width=0.38\textwidth,angle=270]{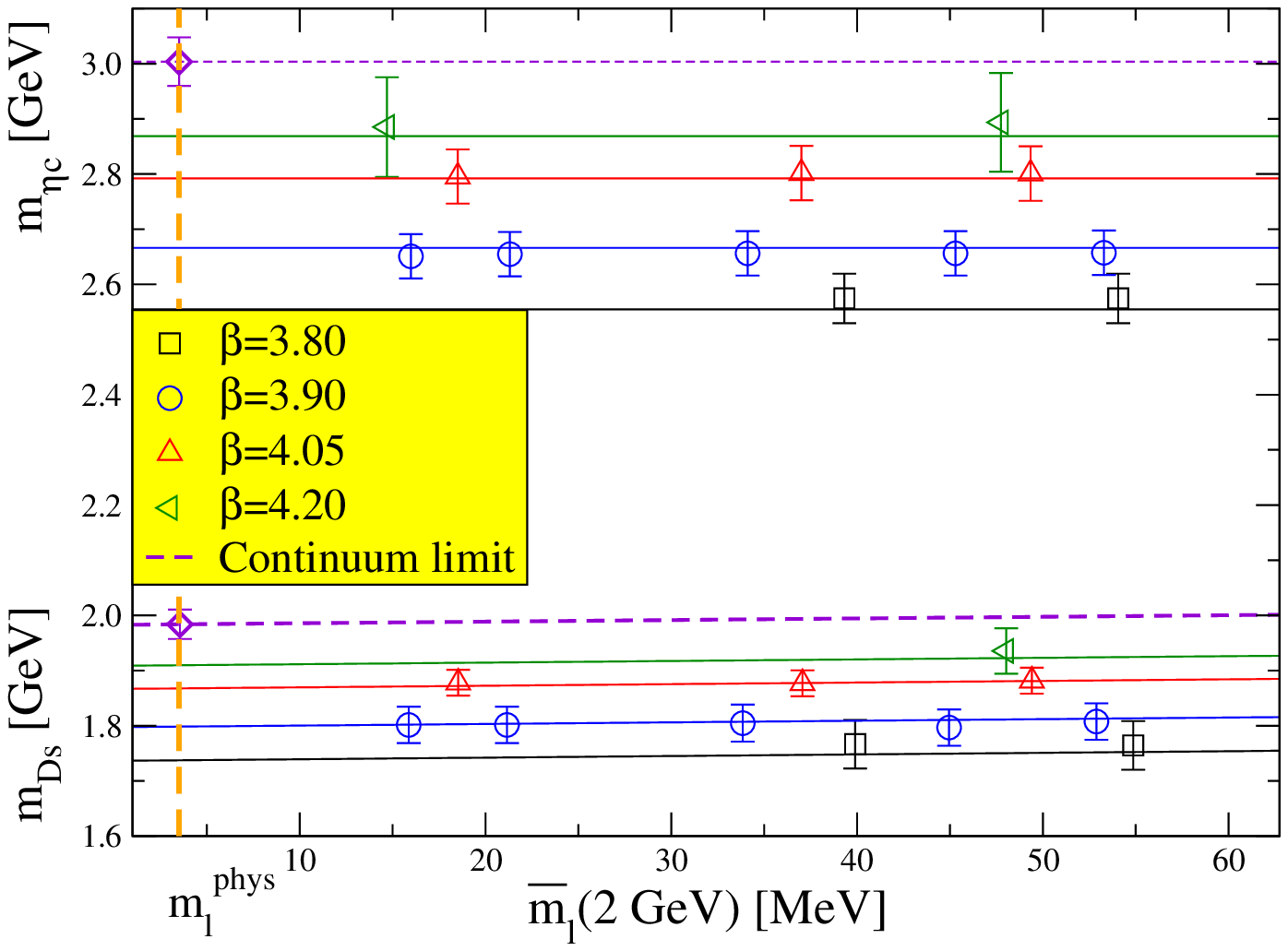} 
\caption{\sl Left: Dependence of $m_D$ (left) and  $m_{D_s}$ and $m_{\eta_c}$ (right) on the light quark mass, at
fixed reference charm quark mass ($\mbar_c^{ref}=1.16\,\gev$) and for the four simulated
lattice spacings.  For the $D_s$ meson the strange quark mass is fixed to the reference value $\mbar_s^{ref}=95 \,\mev$.
\label{fig:mDmDsmetac_vs_ml}}
\end{figure}
\begin{figure}[tb]
\includegraphics[width=0.38\textwidth,angle=270]{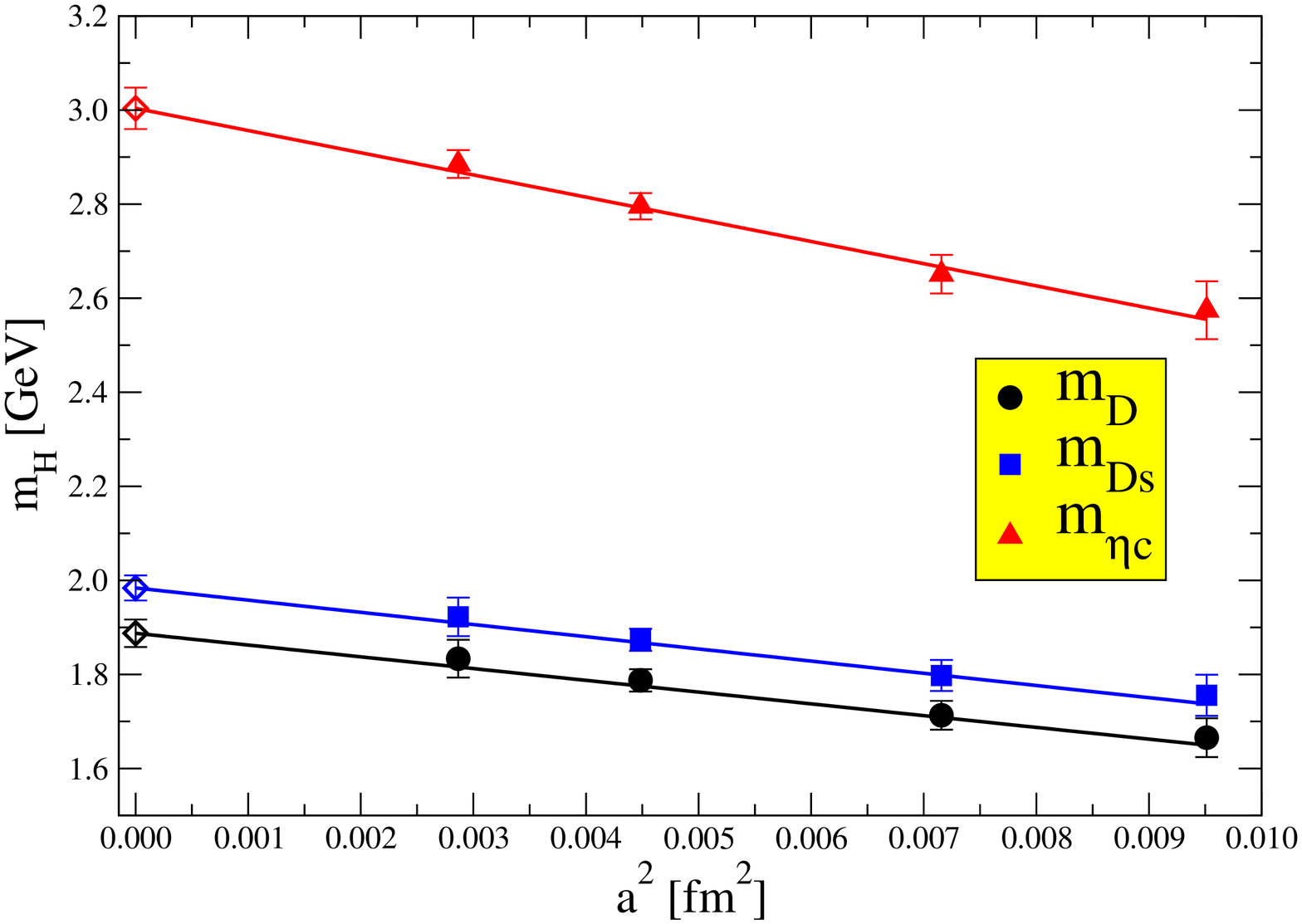}
\hspace{-0.8cm}
\includegraphics[width=0.38\textwidth,angle=270]{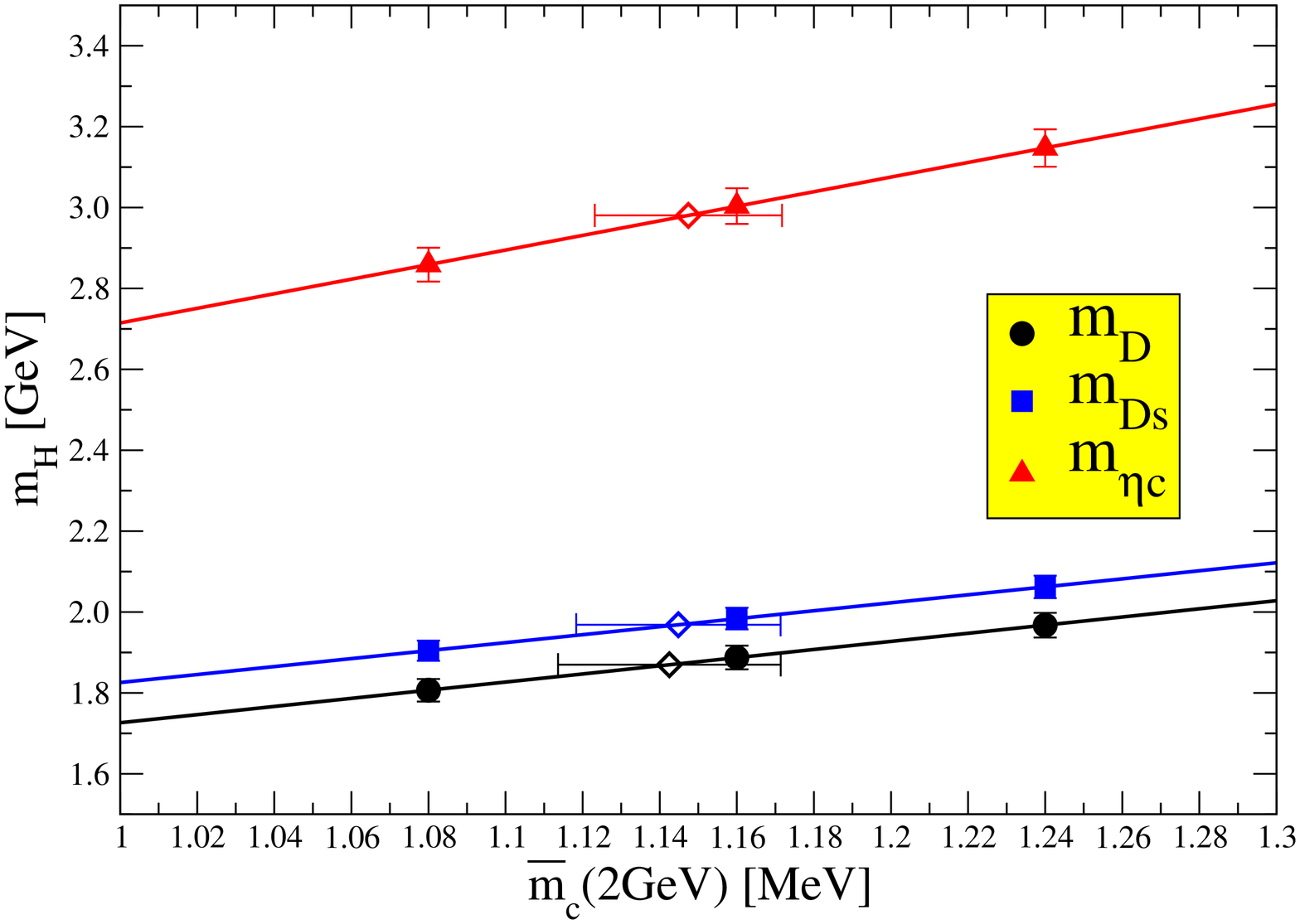} 
\caption{\sl Left: Dependence of $m_D$, $m_{D_s}$ and $m_{\eta_c}$, at
fixed reference charm quark mass ($\mbar_c^{ref}=1.16\,\gev$) and at physical up/down and strange quark mass, on the squared lattice spacing. Right: Dependence of $m_D$, $m_{D_s}$ and $m_{\eta_c}$, in the continuum limit and at physical up/down and strange quarks, on the charm quark mass. The charm mass results from the three determinations are also shown (empty diamonds).
\label{fig:mDmDsmetac_vs_a2mc}}
\end{figure}
As for the strange quark case, we have first used a quadratic spline fit to interpolate the data at three reference values of the charm mass equal at the four $\beta$ values: $\mbar_c^{ref}(2\gev) = 
\{ 1.08\,,\ 1.16\,,\ 1.24\} \gev$. In order to fit the meson masses we have considered (phenomenological) polynomial fits, which turn out
to describe well the dependence on the light and strange quark masses and on the lattice cutoff of the $D$, $D_s$ and
$\eta_c$ meson masses, at fixed (reference) charm mass $m_c$.
Then, the value of the physical charm quark
mass has been extracted by fitting these data as a function of the charm quark mass
and using as an input the experimental value of the corresponding charmed meson
$m_D^{exp}= 1.870\,\gev$, $m_{D_s}^{exp}=1.969\,\gev$, $m_{\eta_c}^{exp}=2.981\,\gev$.
For the charm mass dependence, a constant plus either a linear or a $1/m_c$ term have been considered for describing data of the $D$, $D_s$ and $\eta_c$ mesons. Both choices are found to describe very well the lattice data.
In fig.~\ref{fig:mDmDsmetac_vs_ml} we show the dependence of the $D$, $D_s$ and $\eta_c$ masses on the light quark mass at a fixed reference charm mass, for the four $\beta$'s. For the $D_s$ and $\eta_c$ mesons, which contain the light
quark in the sea only, this dependence is not significant within the statistical errors.
In fig.~\ref{fig:mDmDsmetac_vs_a2mc} (left) the meson masses at  physical light and strange quark masses are shown as a function of $a^2$, for a reference value of the charm quark mass.
Finally, fig.~\ref{fig:mDmDsmetac_vs_a2mc} (right) shows the dependence of the $D$, $D_s$ and $\eta_c$ masses on the charm mass and the interpolation to the physical charm.

In order to evaluate the systematic uncertainty, we have summed in quadrature the approximately 1\% spread among the three determinations from the $D$, $D_s$ and $\eta_c$ mesons, the $1.5$\% uncertainty due to discretization effects (estimated by excluding the data from the coarser lattice) and the 2\% uncertainty coming from the perturbative conversion of the renormalization constants from the RI-MOM to the $\msb$ scheme.
We quote as our final result for the charm quark mass in the $\msb$ scheme
\be
\label{eq:rismc}
\mbar_c(2 \gev)=1.14(3)(3) \gev=1.14(4) \gev\,\qquad
\rightarrow ~~ \mbar_c(\mbar_c)=1.28(4) \gev\,,
\ee
where the evolution to the more conventional scale given by
$\mbar_c$ itself has been performed at N$^3$LO~\cite{Chetyrkin:1999pq} with $N_f=2$, consistently with our non-perturbative evaluation of the renormalization constant.
Our result is in good agreement with the HPQCD result $\mbar_c(\mbar_c)=1.268(9)\,\gev$~\cite{Allison:2008xk}, with a larger uncertainty in our determination, and with the recent sum rules determination $\mbar_c(\mbar_c)=1.279(13)\,\gev$ of~\cite{Chetyrkin:2009fv}.
We have also provided a prediction for the scheme and scale independent ratio
\be
m_c/m_{s}=12.0(3) \ .
\ee

\section{Bottom quark mass}
In these proceedings we also take the opportunity to provide an updated value for the b-quark mass following the method presented in~\cite{Blossier:2009hg}. This method consists in the calculation of the b-quark mass using suitable ratios of the heavy-light pseudoscalar meson masses which, by construction, have an exactly known infinite mass limit.
In the present study we have employed simulation data at three $\beta$ values as in~\cite{Blossier:2009hg}, namely 3.80, 3.90 and 4.05, but using now the final value of the quark mass renormalization constants~\cite{Constantinou:2010gr} and the same statistics as for the other quark mass determinations. In this way we have obtained the updated estimate 
\be
\bar{m}_b(\bar{m}_b)=4.3(2)\,\gev \  .
\ee
This value is about $1\sigma$ smaller, with an improved accuracy with respect to the result given in~\cite{Blossier:2009hg}.


\end{document}